\documentclass[prl,twocolumn,showpacs]{revtex4}
\usepackage{graphicx}
\usepackage{epsfig}
\usepackage{dcolumn}
\usepackage{bm}
\usepackage{amsmath}
\usepackage{amsbsy}
\usepackage{amssymb}
\usepackage{lscape}
\usepackage[usenames]{color}

\newcommand{\bra}[1]{\ensuremath{\langle{#1}|\,}}
\newcommand{\ket}[1]{\ensuremath{\,|{#1}\rangle}}

\begin{document}


\title{Below-bandgap excitation of bulk semiconductors by twisted light}

\author{G.\ F.\ Quinteiro}
\affiliation{ \mbox{CONICET and Departamento de F\'{\i}sica ``J.\
J.\ Giambiagi'', Universidad de Buenos Aires} \mbox{Ciudad
Universitaria, Pabell\'on I, C1428EGA Ciudad de Buenos Aires,
Argentina}}

\date{\today}

\begin{abstract}
I theoretically investigate the response of bulk semiconductors to
excitation by twisted light below the energy bandgap.
To this end, I modify a well-known model of light-semiconductor
interaction to account for the conservation of the light's momentum.
I show that the excited states can be thought of as a superposition
of slightly perturbed exciton states undergoing a complex
center-of-mass motion.
In addition, the absorption would occur at a slightly shifted energy
(compared to plane waves) and would exhibit complex spatial patterns
in the polarization and current.
\end{abstract}

\maketitle



FINAL VERSION PUBLISHED IN EUROPHYSICS LETTERS EPL 91, 27002
(2010) \\

In the realm of semiconductor optics, a generalized practice is to
neglect the photon's momentum in direct absorption/emission
processes. This assumption, known as ``vertical
transitions''(VT)\cite{Haug-01}, has been applied with such success
that sometimes the need to include the momentum is understimated.
Exceptions are found in early and recent works, reporting
theoretical and experimental results strongly dependending on the
conservation of the linear momentum of the photon. In the late '50
Hopfield elaborated the theory of
exciton-polaritons\cite{Hopfield-01}, and later he pointed out, in a
different context, the existence of new selection rules not
accounted for by VT\cite{Hopfield-02}. Afterwards the photon drag
effect\cite{Gibson-01} was proposed, experimentally verified and
brought to the point of technological applications. More recently,
the photon momentum has been incorporated in the equations of motion
describing the dynamics of electrons and holes under the action of
inhomogeneous light fields to treat lasers\cite{Kuhn-01}; meanwhile,
research in exciton-polaritons continue
evolving\cite{Polaritons-01}.

In the last few decades there has been an increase in the number of
studies on spatially inhomogeneous light fields, most important to
this article is the case of optical vortices\cite{OpVortex-01}, and
subsequently that of twisted light(TL) ---light carrying orbital
angular momentum(OAM)\cite{Allen-01}. These have motivated studies
of inhomogeneous light-matter interaction in several
fields\cite{LightMattI-01}, and likely part of the work in
semiconductors cited above. In particular, theoretical work by
myself and collaborators show the need to include the OAM of TL in
above-bandgap excitation of semiconductor
structures.\cite{Quinteiro-1, Quinteiro-2, Quinteiro-3}

From the point of view of applications, the tendency to
miniaturization requires the manipulation of states having few
particles, and the consideration of small quantities. Thus, what has
been judgicioulsly neglected in the past, may acquire relevance in
present-days technology.

The present work is motivated by the combination of the current
interest in optical vortices, the long-standing attention to the
light-matter interaction problem, and finally the intention to gain
control on the quantum level in semiconductor structures. In the
following I explain the modifications introduced in a standard model
of light-semiconductors interaction, to account for the linear
momentum or OAM of the light.
Using this extended model I work out the familiar situation of
excitation of excitons by plane waves. I have in mind two aims;
first, I intend to further support ---in addition to the
aforemention literature--- the idea that the inclusion of the
momentum of light is a must, for certain situations. Second, some
simple new results and connection to other theories will emerge.
The main part of the article treats the below-bandgap coherent
excitation by TL.

The theory of optical excitation of semiconductors in the VT limit
is well developed and known. An excellent treatment is given by Haug
et al\cite{Haug-01}, for the cases without (inter-band transitions)
and with  (excitonic transitions) Coulomb interaction. In the
following, this formalism will be extended to include the momentum
of light.

The dynamics of electrons in a semiconductor having a valence($v$)
and a conduction($c$) bands may be described by Heisenberg equations
of motion\cite{Quinteiro-3} for populations in each band plus the
quantum coherence between them; these equations form a coupled
system. Under the condition of low excitation (low light field
intensity/large detuning), the equations can be treated
perturbatively and can be decoupled.
\footnote{Perturbation theory breaks down when the quantization
volumes for electrons and photons is the same and no additional
dissipation channel exists.\cite{Hopfield-01} On the other hand, the
conservation of momentum requires that the system is larger than the
wavelength $\lambda$ of light; thus, for a system smaller that the
photon's quantization box and larger than $\lambda$ perturbation
theory can be used.}
The first order term in the light field is the quantum interband
coherence $\rho_{vi,cj}= \langle a^\dag_{vi} a_{cj} \rangle$, where
$a^\dag_{bk}/a_{bk}$ are creation/annihilation operators of
electrons in a Bloch state $\mathbf k$ and band $b$: $\psi_{b\mathbf
k}(\mathbf r)= \bra{\mathbf r}{b \mathbf k}\rangle= e^{i\mathbf k
\mathbf r} u_b(\mathbf r)$, with energy $\varepsilon_{b \mathbf{k}
}$, and periodic Bloch function $u_b(\mathbf r)$; whereas,
populations are second order in the light field. Then, the linear
response of the unexcited system (zero conduction band population)
is obtained from
\begin{eqnarray} \label{Eq:Interband_Pol}
    \left( i\hbar \frac{d}{dt}
          - \Delta_{c \mathbf{k}, v \mathbf{k}' }
    \right) \rho_{v \mathbf{k}', c \mathbf{k}} (t)
&=& \bra{c \mathbf{k}} h_I(t) \ket{v \mathbf{k}'} - \nonumber \\
&&  \hspace{-10mm}
    \sum_{\mathbf{q} \neq 0} V_{\mathbf{q}} \,
    \rho_{v \mathbf{k}'-\mathbf{q},c \mathbf{k}-\mathbf{q}}(t)
\,,
\end{eqnarray}
where $\Delta_{c \mathbf{k}, v \mathbf{k}' } = \varepsilon_{c
\mathbf{k} } - \varepsilon_{v \mathbf{k}' }$ and the self-energy
correction has been neglected. The first term in the RHS is the
matrix element of the light-matter interaction, modeled by ``minimal
coupling'' (for the case of TL see Quinteiro et al
\cite{Quinteiro-1}). The second term on the RHS is the Coulomb
potential in the random phase approximation. I first note that, due
to the conservation of the momentum of light, $\bra{c \mathbf{k}}
h_I(t) \ket{v \mathbf{k}'}$ vanishes unless $\mathbf{k}' =
\mathbf{k}-\mathbf q_0$ for some vector $\mathbf q_0$.
In the case of plane waves propagating along the $z$-axis $\mathbf
q_0 = q_0 \hat{z}$ is a constant or parameter ---the linear momentum
of the photon---, while for twisted light $\mathbf q_0 = q_\parallel
\cos \theta \,\hat{x} + q_\parallel \sin \theta \,\hat{y} + q_z
\,\hat{z}$ with $\theta$ a variable and $\{q_\parallel, q_z\}$
parameters (see below for details). Therefore, to capture the
relevant physics of the light-matter interaction I specialize and
study Eq.\ (\ref{Eq:Interband_Pol}) at $\mathbf{k}' = \mathbf{k} -
\mathbf q_0$.
I will treat the case of TL, since that of plane waves can be easily
deduced from the former. The main contribution to the light-matter
interaction arises from the transverse component
$\mathbf{A}(\textbf{r},t) = \boldsymbol{\epsilon}_\sigma \, A_0(t)
J_l(q_\parallel\, r_\parallel)\, \exp[{i (q_z z + l\phi)}] + c.c$ of
the vector potential, with $J_l(x)$ a Bessel function, $\hbar l$ the
OAM,  $\boldsymbol\epsilon _\sigma$ the vector for circular
polarization $\sigma=+/-$, and $q_\parallel < q_z$.
 A Fourier transform in $\mathbf k$-space $\{f(\mathbf{r})=[L/(2\pi)]^3
\, \int \, d^3k \,e^{-i \mathbf{k} \mathbf{r}} f(\mathbf{k})\}$ and
time [$f(t)=1/(2\pi) \, \int \, d\omega\, e^{-i \omega t}
f(\omega)$] is applied to simplify each term: The
light-matter interaction  $\bra{c \mathbf{k}} h_I(t) \ket{v
\mathbf{k}'} = \xi(t)\, \delta_{{\boldsymbol \kappa}_\parallel
q_\parallel} \delta_{{\boldsymbol \kappa}_z q_z}\exp{(i\theta
l)}/(L\,q_\parallel)$ becomes
%
\begin{eqnarray*}
    \bra{c \mathbf{k}} h_I(t) \ket{v \mathbf{k}-\mathbf q_0 }
&\rightarrow&
   L^3 \, \xi(\omega)  \, \frac{e^{i \theta l}}{L\, q_\parallel} \, \delta(\mathbf{r})
\end{eqnarray*}
with $\xi(t) = -(-i)^l \,(\boldsymbol{\epsilon} \cdot
\mathbf{p}_{cv})\, Q\,A_0(t)/m$, $\mathbf{p}_{cv} = \bra{u_c}\mathbf
p \ket{u_c}$ the momentum-operator matrix element, $Q$ the
electron's charge, $m$ the electron's mass, and $L$ the linear size
of the system. Then,
%
\begin{eqnarray}\label{Eq:F_DeltaRho}
    \Delta_{c \mathbf{k}, v \mathbf{k} - \mathbf q_0 }
    \,\rho_{v \mathbf{k} - \mathbf q_0 , c \mathbf{k}} (\omega)
&\rightarrow& \left[
        \left(E_g + \frac{\hbar^2\, \mathbf q_0^2}{2|m_v^*|}\right) -
    \right. \nonumber \\
&&  \hspace{-20mm}
    \left.
        i\, \frac{\hbar^2\,\mathbf q_0}{|m_v^*|} \cdot \nabla -
        \frac{\hbar^2}{2\, \mu} \nabla^2
    \right]
    \, \rho_{\mathbf q_0}(\omega, \mathbf{r})
\end{eqnarray}
with $1/ \mu = 1/|m_v^*| + 1/|m_c^*|$, and the subscript $v/c$ from
$\rho$ was eliminated to ease the notation. Finally,
%
\begin{eqnarray} \label{}
    \frac{L^3}{(2\,\pi)^3} \int d^3j\, V_{\mathbf{k} - \mathbf{j}} \,
    \rho_{v \mathbf{j} - \mathbf q_0 ,c \mathbf{j}}(\omega)
    \rightarrow
    V(\mathbf{r})\, \rho_{\mathbf q_0} (\omega, \mathbf{r})
\,.
\end{eqnarray}
Assembling all terms, the transformed version of Eq.
(\ref{Eq:Interband_Pol}) becomes
\begin{eqnarray} \label{Eq:Interband_Pol1.1}
&&  \left[
        \hbar \omega -
        \left(E_g + \frac{\hbar^2\,\mathbf q_0^2}{2|m_v^*|}\right) +
        i\, \frac{\hbar^2\,\mathbf q_0}{|m_v^*|} \cdot \nabla +
        \frac{\hbar^2}{2\, \mu} \nabla^2 +
        \right. \nonumber \\
&&  \hspace{5mm}
        V(\mathbf{r})
     \Big] \rho_{\mathbf q_0 } (\omega, \mathbf{r})
=   L^3
    \, \xi(\omega)\,\frac{1}{L\,q_\parallel} e ^{i\theta l} \,
    \delta(\mathbf{r})
\,
\end{eqnarray}
A comparison with the VT model ($\mathbf q_0 = 0$) reveals two new
terms: $(1)$ $\hbar^2\,\mathbf q_0^2/(2|m_v^*|)$, and $(2)$ $i
\hbar^2/|m_v^*| \,\mathbf q_0 \cdot \nabla$. Term $(1)$ needs no
special care, since it is a renormalization of the energy.  Term
$(2)$ can be handled by either $i)$ perturbation theory, or $ii)$
completing squares and applying a unitary transformation
(displacement in $\mathbf k$-space space) $U(\mathbf r) =
\exp\left[i (\mu/|m_v^*|) \,\mathbf q_0 \cdot \mathbf r \right]$. I
will mainly follow method $ii)$; however, the perturbation theory
analysis yields useful information. First, I show that the
perturbation is indeed small compared to the rest. The solutions to
the unperturbed homogeneous equation [RHS of Eq.
(\ref{Eq:Interband_Pol1.1})] are the excitonic relative-motion
wave-functions $\psi_\nu(\mathbf r)$ with energy $E_\nu$ and spatial
extent $a_B^*$; then, $\hbar^2/|m_v^*| \,\mathbf q_0 \cdot \nabla
\psi_\nu(\mathbf{r}) \simeq \hbar^2 q_0/(a_B^* |m_v^*|)  \,
\psi_\nu(\mathbf{r})$. Typically (e.g. GaAs)  $a_B^* \simeq 10$ nm
and $q_0 \simeq 10^{-2}$ nm$^{-1}$, so $\hbar^2 q_0/(a_B^* |m_v^*|)
\simeq 0.1$ meV which is smaller than $ E_\nu \simeq 1$ meV, as
required. The ratio $(2)/(1)\simeq 8$, so I retain both terms. By
method $ii)$ Eq. (\ref{Eq:Interband_Pol1.1}) becomes
\begin{eqnarray} \label{Eq:Interband_Pol2}
    \left[
        \hbar\,\omega - E_g  -
        \frac{\hbar^2 \mathbf q_0^2}{2 M} -
        \frac{\mathbf p^2}{2\mu} +
        V(\mathbf{r})
    \right]
    \widetilde{\rho}_{\mathbf q_0} (\omega, \mathbf{r}) = \nonumber \\
&&  \hspace{-45mm}
    L^3
    \, \xi(\omega) \, \frac{e^{i \theta l}}{L\, q_\parallel} \, U(\mathbf r) \,\delta(\mathbf{r}) \,,
\end{eqnarray}
where $\widetilde{\rho}_{\mathbf q_0} (\omega, \mathbf{r}) =
U(\mathbf r) {\rho}_{\mathbf q_0} (\omega, \mathbf{r})$, and $M =
|m^*_v| + |m^*_c|$. Redefining $E_g' = E_g + \hbar^2 \mathbf
q_0^2/(2 M)$ the homogeneous part becomes the equation for the
relative motion of an exciton. Back to $\mathbf k$-space, the
complete solution is
\footnote{To account phenomenologically for decoherence $\omega$ may
be replaced by $\omega +i\gamma$ in the denominator.}
\begin{eqnarray}\label{Eq_Coherence_TS_final}
    \rho_{v \mathbf{k} - \mathbf q_0 ,c \mathbf{k}}(\omega)
&=& L^3\, \xi(\omega)\,
    \frac{e^{i \theta l}}{L\, q_\parallel}  \times \nonumber \\
&&  \hspace{-20mm}
    \sum_\nu \,
    \frac{\psi^*_\nu(\mathbf{r}=0)}
    {\hbar\,\omega - E_g - \frac{\hbar^2 \mathbf q_0^2}{2 M} - E_\nu}
    \psi_\nu \left(\mathbf{k} - \frac{\mu}{|m_v^*|} \mathbf q_0 \right) \,.
\end{eqnarray}
Later, it will prove useful a Taylor expansion of $\psi_\nu
(\mathbf{k} - \mu \mathbf q_0/ |m_v^*| )$, which is possible since
each term is of the order of $\psi_\nu (\mathbf{k}) (q_0 a_B^* \mu /
|m_v^*|)^n$, with $q_0 a_B^* \mu / |m_v^*| < 1$.  Notice the
correspondence between this expansion and the perturbation theory
approach $i)$.

Equation (\ref{Eq_Coherence_TS_final}) is the building-block for
several quantities describing both the electrons' kinetics/dynamics
and the effect that electrons have on the EM-field.
Next, I provide the expectation values of the polarization, the
electric current, and the OAM.
\footnote{These quantities were obtained from expressions in terms
of field operators, and though straightforward, the derivation is
lengthy and so not given.}

As a consequence of the conservation of photon momentum, the global
polarization of the system is zero; thus, a more correct quantity is
a local or space-dependent polarization
\begin{eqnarray}\label{Eq_Pol}
    {\mathbf P}({\mathbf R}, t) \! \!
&=& \!  \! 2 \left(\frac{L_{\cal R}}{L}\right)^{\!3}
    \! \sum_{\mathbf k \mathbf q_0 }
    \Re\left\{e^{i \mathbf q_0 \cdot \mathbf R}
    \,\mathbf d_{vc} \,
    \rho_{v \mathbf k - \mathbf q_0, c \mathbf k }(t)\right\}
    \!,
\end{eqnarray}
with $\Re\{\ldots\}$ the real part, $\mathbf d_{vc}$ the dipole
matrix element, and $\mathbf R$ pointing to a macroscopic cell of
linear size $L_{\mathbf R}$ small compared to the scale of variation
of the EM-field, but larger than the unit cell of the semiconductor.
In addition, a TL-field induces electric currents\cite{Quinteiro-1,
Quinteiro-2}
\begin{eqnarray}\label{Eq_Cur}
    \mathbf j({\mathbf R}, t)
&=& 2 \frac{Q}{m}
    \frac{1}{L^3} \sum_{\mathbf k \mathbf q_0}
    \Re\left\{e^{i \mathbf q_0 \cdot \mathbf R}
    \, \mathbf p_{vc} \,
    \rho_{v \mathbf k - \mathbf q_0, c \mathbf k }(t)\right\}
    \,.
\end{eqnarray}
The similarity between the polarization and current is not
surprising; the theory of macroscopic media relates the polarization
charge to the electric current by the continuity equation. Either
one or the other may be used to study the effect that the interband
coherence Eq.\ (\ref{Eq_Coherence_TS_final}) has on the EM-field.
It is also meaningful to calculate to OAM acquired by the electrons.
For reasons similar to those that cause the global polarization to
vanish, only an OAM in slices perpendicular to the propagation
direction of light yields a non-zero result
\begin{eqnarray}\label{Eq_Lz}
    L_z(Z, t)
&=& \beta \,e^{i q_z Z}\, \sum_{\mathbf{k} \mathbf q_0 }
    \Re\left\{
    ( p_{-,b'b} \,e^{i \theta}  + p_{+,b'b} \,e^{-i \theta} ) \times
    \right. \nonumber \\
&&
    \left.
    \,\rho_{v \mathbf{k} -\mathbf q_0, c \mathbf{k}}(t)\right\}
\,,
\end{eqnarray}
where $ \beta =- (4\pi/ q_\parallel)\, J_2(q_\parallel L) $, and
$p_{\pm,b'b} = p_{x,b'b} \pm p_{y,b'b}$.
Some general comments are in order: $i)$ Because of their local
character, all quantities exhibit spatial dependence; $ii)$ to
obtain explicit expressions, it only remains to insert the
time-domain version of $\rho_{v \mathbf{k} - \mathbf q_0 ,c
\mathbf{k}}(\omega)$ into Eqs. (\ref{Eq_Pol})-(\ref{Eq_Lz}), or to
transform all quantities to the frequency domain and use $\rho_{v
\mathbf{k} - \mathbf q_0 ,c \mathbf{k}}(\omega)$ directly; $iii)$
for plane waves all sums over $\{\mathbf{k}, \mathbf q_0\}$ simplify
to $\{\mathbf{k}\}$, while for TL they simplify to $\{\mathbf{k},
\theta \}$.

With the tools developed so far, I analyze the cases of plane waves
and twisted light.

\textit{Plane waves:} I set $l=0$ and $e^{i \theta l}/(L\, q_\parallel) = 1$.
The simplest situation is that of $q_0 = 0$, i.e. VT. By noting that
$\sum_k \psi_\nu(\mathbf k) \propto \psi_\nu(\mathbf r = 0)$ the
expressions for the coherence and the polarization boil down to
those of the standard result, that is the polarization is
homogeneous and new absorption lines below the conduction-band edge
appear due to the exciton's binding energy $E_\nu$.
Let us now turn to the general case $\mathbf q_0 = q_0 \hat{z}$, and
examine Eq. (\ref{Eq_Coherence_TS_final}) after expanding $\psi_\nu
(\mathbf{k} - \mu \mathbf q_0/ |m_v^*| ) =  \psi_\nu (\mathbf{k}) +
\sum 1/n! (\mu q_0/|m_v^*|)^n \partial_{k_z}^n \psi_\nu (\mathbf{k})
$. The first term is similar to that of VTs; nevertheless, the
presence of the kinetic energy term $\hbar^2  q_0^2/(2 M)$ in the
denominator signals the center-of-mass (COM) motion of the exciton,
as has been known for long time.
The next terms in the expansion of Eq. (\ref{Eq_Coherence_TS_final})
are the consequences of the displacement in momentum space of the
excitonic relative-motion wave-function, as seen by the argument of
$\psi_\nu (\mathbf{k} - \mu \mathbf q_0/ |m_v^*| )$. From the point
of view of perturbation theory, the term $i \hbar^2/|m_v^*|
\,\mathbf q_0 \cdot \nabla$  in Eq. (\ref{Eq:Interband_Pol1.1})
causes the eigenstates to be a superposition of unperturbed
excitonic wave-functions. Both standpoints tell us that this
correction affects the internal degree of freedom, in contrast to
the effect $\hbar^2\,\mathbf q_0^2/(2|m_v^*|)$ has.
Shifting to the analysis of derived quantities, we see that the
polarization presents spatial dependency on the coordinate $Z$; from
this expression an electric susceptibility in the direction of the
electric field can be deduced. Expanding the real part in Eq.
(\ref{Eq_Pol}) and after some calculation
\begin{eqnarray}\label{Eq_Chi}
    \chi_{}( \omega)
&=& \!\!- |{ d}_{v c}|^2
    \,  \sum_\nu \,
    \frac{\psi^*_\nu(\mathbf{r}=0)}
    {\hbar\,\omega - E_g - \frac{\hbar^2 q_0^2}{2 M} - E_\nu}
    \left[
        \psi_\nu\left(\mathbf{r} = 0\right) +
   \right. \nonumber \\
&& \hspace{-8mm}
    \left.
        \sum_{n=1}^\infty
        \frac{1}{n!}
        \left(\frac{\mu q_0}{|m_v^*|}\right)^n
        \left(\frac{L}{2\,\pi}\right)^3\,
        \int_{}\, d^3 \mathbf k \,
        \partial_{k_z}^n
        \psi_\nu \left(\mathbf{k}\right)
    \right] + \ldots
\end{eqnarray}
where the dots indicate an extra term arising from $\left\{e^{i
\mathbf q_0 \cdot \mathbf R}    \,\mathbf d_{vc} \,    \rho_{v
\mathbf k - \mathbf q_0, c \mathbf k }(t)\right\}^*$. The
susceptibility shows the expected features, i.e. no spatial
dependence
\footnote{The $\exp(i q_0 Z)$ was used to reconstruct the electric
field in a formula of the type $P(Z)=\chi E(Z)$.}
but ``spatial dispersion'' ($q_0$ dependence). The COM-motion
correction in the denominator introduces a tiny shift in the
absorption line.
Used in conjunction with Maxwell's equations, the susceptibility
yields information about the effect that electrons has on the EM
field, e.g. attenuation of the beam.
I finally note that the effects introduced by a finite wave-vector
$q_0$ is negligible when the semiconductor's $z$-length is smaller
than $2\pi/q_0$, e.g. thin samples/quantum wells.

\textit{Twisted light:} The full solution Eq.
(\ref{Eq_Coherence_TS_final})  having $e^{i \theta l}/(L\,
q_\parallel) \neq 0$, and variable $\theta$ is used.
Let us first focus on the term $\hbar^2  q_0^2/(2 M)$. Regarded as
the COM kinetic energy of excitons, it indicates a complex motion
connected to both parameters $q_z$ and $q_\parallel$.
Although $q_\parallel$ is not a momentum (it relates to the inverse
of the beam waist), its presence may be understood by thinking on
the linear momentum ($\mathbf P^{(TL)}= \epsilon_0 \mathbf E \times
\mathbf B$) at each point in space carried by the TL field: A
preliminary calculation, under the  assumption $q_\parallel < q_z$,
shows that the radial and angular components of $\mathbf P^{(TL)}$
are proportional to $q_\parallel$, while ---as expected--- its $z$
component is proportional to $q_z$. Despite this tells that
$q_\parallel$ participates in the momentum, a deeper analysis
remains to be done in order to confirm and expand this result.
Once again, the argument of $\psi_\nu (\mathbf{k} - \mu \mathbf q_0/
|m_v^*| )$ indicates that the TL-field excites states which are not
exactly but slightly perturbed excitonic states.
Additionally, the presence of the factor $e^{i \theta l}$ signals
that a superposition of states ---differing in their $\theta$
variable and thus having different phases--- is created, in contrast
to the case of the one-state excitation by plane waves.
As for the macroscopic description of the system, the states
contribute, by the term  $E_\nu + \hbar^2  q_0^2/(2 M)$, absorption
lines below the band gap; the shift due to the COM kinetic energy is
about $10\mu$eV. The absorption could be studied more carefully by
either calculating the local polarization or the electric current,
and if possible deriving a susceptibility or conductivity (through
Ohm's law) respectively.
It is of interest to look at the electric current produced by the
interband coherence; using Eq.\ (\ref{Eq_Coherence_TS_final}) and
defining $ \rho_{v \mathbf k - \mathbf q_0, c \mathbf k }(t)= e^{i
\theta l} \rho_0 (t)$ Eq.\ (\ref{Eq_Cur}) becomes
\begin{eqnarray*}
    \mathbf j(Z, \Phi ; t)
&=& \frac{Q}{m}
    \frac{4\pi }{L^3} \sum_{\mathbf k }
    \Re\left\{ i^l
    e^{i q_z Z }
    J_l(q_\parallel R)
    e^{i l \Phi}
   \mathbf p_{vc} \,
    \rho_{0}(t)\right\}
    \,.
\end{eqnarray*}
The current perpendicular to the $z$-axis exhibits complex flow
patterns: for $l=1$ one observes circular flows around the beam
axis, for $l=2$ two electric current vortices appear at both sides
of the beam axis, for $l>2$ the complexity increases and several
vortices show up (see Fig. 2 in Quinteiro et al \cite{Quinteiro-1}),
suggesting a transfer of the optical vortices to the solid. As a
consequence, the OAM in $z$-direction only exists for the case of
$l=\pm 1$.
To conclude, it is my believe that the the correct physical
description of the TL-semiconductor interaction requires the
inclusion of $\mathbf q_0$; this is simply because common
semiconductor structures are not shaped to avoid the angular
inhomogeneity of a TL beam.
\footnote{An exception would be an arc-section sample, subtending a
small angle, whose origin coincides with the the beam axis.}
%

In conclusion, a modified version of a well-established model has
been presented and used to explain the below-bandgap excitation of
semiconductors by inhomogeneous light beams, with special emphasis
on twisted light.
First, I show how the standard results for the case of vertical
transitions induced by plane waves are obtained  in the appropriate
limit; this attests for the robustness of the extended model.
Next, the case of plane waves when the wave vector of the light beam
is taken into account is analyzed; I show how to recover the
exciton's COM motion, and an additional correction to the
relative-motion wave-function is found.
Then, the theory is applied to the case of twisted light. I predict
that the optical excitation produces a superposition of states that
differ slightly from excitons and undergo complex center-of-mass
motion. In addition, I show that the absorption energy is modified
by the center-of-mass motion, and that the polarization and electric
current induced by the transition present a complex spatial pattern.
Given the current interest in the interaction of inhomogeneous
EM-fields with semiconductors, and in particular the work in
exciton-polaritons, further research on the generation of
exciton-like states and their complex motion due to twisted light
excitation may significantly impact the basic research and
applications in these areas.

I thank P. I. Tamborenea and J. Berakdar for fruitful discussions,
and support by ANPCyT.


\end{document}